\documentclass[reprint,prl,amsmath,amssymb,superscriptaddress]{revtex4-2}
\usepackage[T1]{fontenc}
\usepackage{color}
\usepackage{graphicx}
\usepackage{epstopdf}
\usepackage{bm}
\usepackage{dcolumn}

\def\ba {\mathbf{a}}

\begin{document}
\title{Universal properties of anyon braiding on one-dimensional wire networks}
\author{Tomasz Maci\k{a}\.{z}ek}
\email[]{kk19347@bristol.ac.uk}
\affiliation{School of Mathematics, University of Bristol, Fry Building, Woodland Road, Bristol BS8 1UG, UK}
\affiliation{Center for Theoretical Physics, Polish Academy of Sciences, Al. Lotnik\'ow 32/46, 02-668 Warszawa, Poland}
\author{Byung Hee An}
\affiliation{Department of Mathematics Education, Kyungpook National University, Daegu, South Korea}

\begin{abstract}
We demonstrate that anyons on wire networks have fundamentally different braiding properties than anyons in 2D. Our analysis reveals an unexpectedly wide variety of possible non-abelian braiding behaviours on networks. The character of braiding depends on the topological invariant called the connectedness of the network. As one of our most striking consequences, particles on modular networks can change their statistical properties when moving between different modules. However, sufficiently highly connected networks already reproduce braiding properties of 2D systems. Our analysis is fully topological and independent on the physical model of anyons. 
\end{abstract}

\maketitle
{\it Introduction.} -- Studies of anyon braiding on one-dimensional wire networks are at the forefront of research of architectures for topological quantum computers. Such a computer would perform its tasks using topological states of matter (describing anyons) that are intrinsically robust against different types of noise and decoherence \cite{reviews}. Anyons arise in quantum systems that are effectively one- or two-dimensional. Braiding of anyons transforms a state of the corresponding quantum system by a unitary operator which is a topological quantum gate. A robust realisation of controlled braiding of anyons is one of the major challenges in this field. Recently developed experimental and theoretical proposals address this challenge by exploring the possibility of braiding of anyons on junctions of one-dimensional wire networks \cite{alicea,alicea-rev}. Such networks are believed to provide a platform for engineering anyonic braiding most easily.

This paper shows that braiding of anyons on networks provides a wider range of possibilities for the resulting topological quantum operations in comparison to 2D architectures. This suggests that there may exist quantum systems where computational universality can be accomplished more easily than in currently known proposals. Our paper also provides a mathematical justification for a widely assumed fact that braiding rules in 2D are compatible with braiding rules on 1D networks. Our purpose here is to describe the above new results, whose mathematical details will be spelled elsewhere \cite{BM}. 

Of particular importance in this context is Kitaev's superconducting chain that supports Majorana edge modes. Such a chain can be modelled as semiconductor nanowires coupled to superconductors \cite{alicea} as well as in other solid state \cite{al-in-as,mourik,rokhison,perge} and photonic systems \cite{pachos}. Braiding of edge modes is then realised by coupling endpoints of wires so that they form a network or, in the simplest case, a trijunction \cite{alicea}. Importantly, Majorana edge modes braid in a non-abelian way making them useful for quantum computation. However, the set of gates obtained by braiding of Majorana fermions is never universal and in order to realise universal quantum computation one has to pursue certain additional strategies \cite{flux,rozek}. This proposal has been recognised as one of the most robust candidates for an architecture of a topological quantum computer. Experimental proposals of the above mentioned trijunction have been made so far including photonic systems \cite{pachos} and Josephson junctions \cite{josephson1,josephson2}. We also mention in this context effective hopping models for anyons that have been studied in \cite{HKR} and that have led to the classification of abelian quantum statistics on networks \cite{HKRS}.

\begin{figure}[h]
\centering
\includegraphics[width=0.5\textwidth]{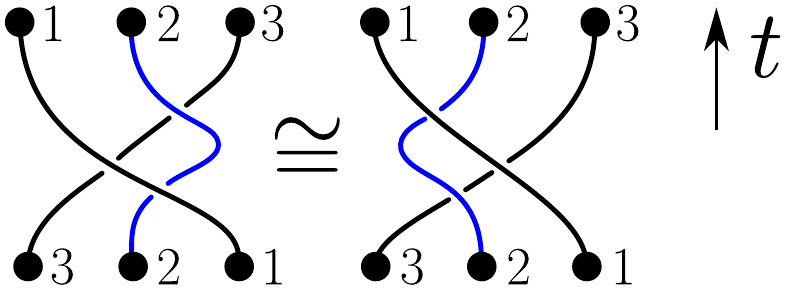}
\caption{The braiding relation for anyons in 2D.}
\label{braid}
\end{figure}

\begin{figure}[h]
\centering
\includegraphics[width=0.5\textwidth]{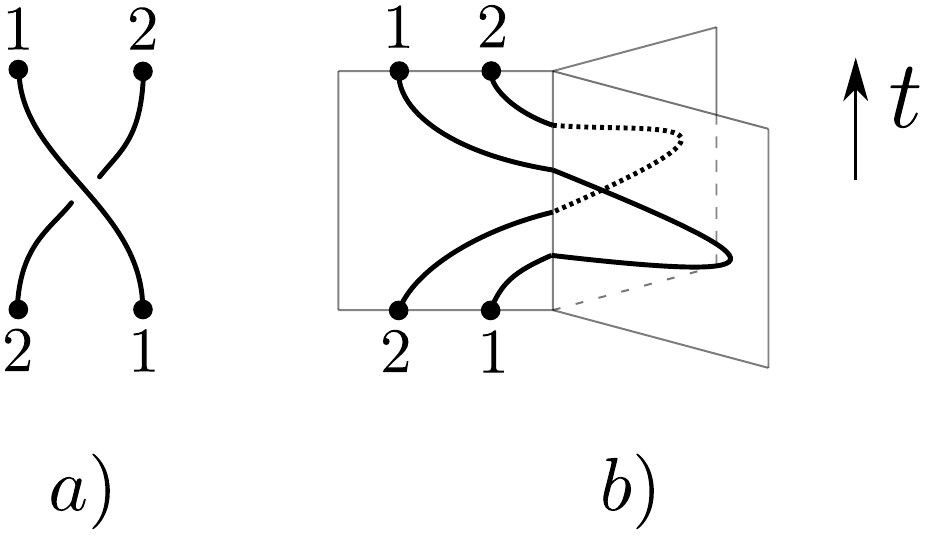}
\caption{Simple braid on the plane vs. simple braid on trijunction.}
\label{comparison}
\end{figure}

Despite the significant interest in problems related to braiding of anyons on networks, relatively little is known about their topological braiding properties. In this work, we fill this gap by studying relations coming from continuous deformations of paths corresponding to braiding of anyons on a network. Because quantum statistics is a topological property, any physical model that supports anyonic braiding on a network has to respect such relations. In other words, any topological transformation of the quantum system related to an exchange of anyons remains invariant under a continuous deformation of the corresponding braid \cite{LM,Souriau,Einarsson}. In the standard 2D setting, an example of such a relation is shown on Fig.\ref{braid}. It relates two ways of exchanging a triple of anyons. To see this, consider firstly the so-called simple braid from Fig.\ref{comparison}a that exchanges two neighbouring anyons. Fig.\ref{comparison}a also explains the origin of term {\it braiding} as the world lines of anyons form braids in space-time.
Let us denote such a simple braid that exchanges $i$th and $(i+1)$th anyon by $\sigma_i$. Any exchange of anyons in 2D can be written as a composition of simple braids. More formally, simple braids generate the planar braid group. However, they do not generate the braid group freely, as they are subject to the following braid relation: $\sigma_i\sigma_{i+1}\sigma_i=\sigma_{i+1}\sigma_i\sigma_{i+1}$. This can be seen on Fig. \ref{braid} -- both braids differ by a deformation of the middle world line (blue line on Fig. \ref{braid}).

Simple braids also satisfy a commutative relation where exchanges of disjoint sets of anyons commute with each other, $\sigma_i\sigma_j=\sigma_j\sigma_i$ for $|j-i|\geq 2$. 

{\it Braiding on junctions.} -- In order to see if the braid relation is satisfied by braids on the trijunction, we first define the network counterpart of the simple braid. It is shown on Fig. \ref{comparison}b -- the trijunction stretched in a time interval makes up three rectangles. Anyon $1$ is first transported to the right branch (the bottom branch in the picture) of the trijunction, then anyon $2$ travels to the left branch making space for anyon $1$ to go back to the original initial position of anyon $2$. The exchange is completed by the return of anyon $2$ from the right branch to the original initial position of anyon $1$. In order to track all moves of anyons on an arbitrary junction, we set up the following notation. For a $d$-junction ($d$ incident branches), we fix the initial position of anyons to align one after another on a fixed branch. Having drawn the junction on a plane, we enumerate the remaining branches in a clockwise fashion by labels from $1$ to $d-1$. The exchange of $i$th and $(i+1)$th anyon will be unambiguously encoded by a sequence of integers $\ba:=(a_1,a_2,\hdots,a_{i+1})$ with $1\leq a_j\leq d-1$ and $a_i\neq a_{i+1}$. Elements of $\ba$ denote i) labels of branches where first $(i-1)$ anyons were distributed -- these are $a_1,\hdots,a_{i-1}$ ii) labels of branches where anyon $i$ and $(i+1)$ exchange -- these are $a_i$ and $a_{i+1}$. Note, that swapping the order of $a_i$ and $a_{i+1}$ reverses the direction of the exchange. Going back to the concrete example of two particles on a trijunction from Fig.\ref{comparison}b, the depicted braid would be denoted by $\sigma_1^{(2,1)}$, i.e. $\ba=(2,1)$.

\begin{figure}[h]
\centering
\includegraphics[width=0.5\textwidth]{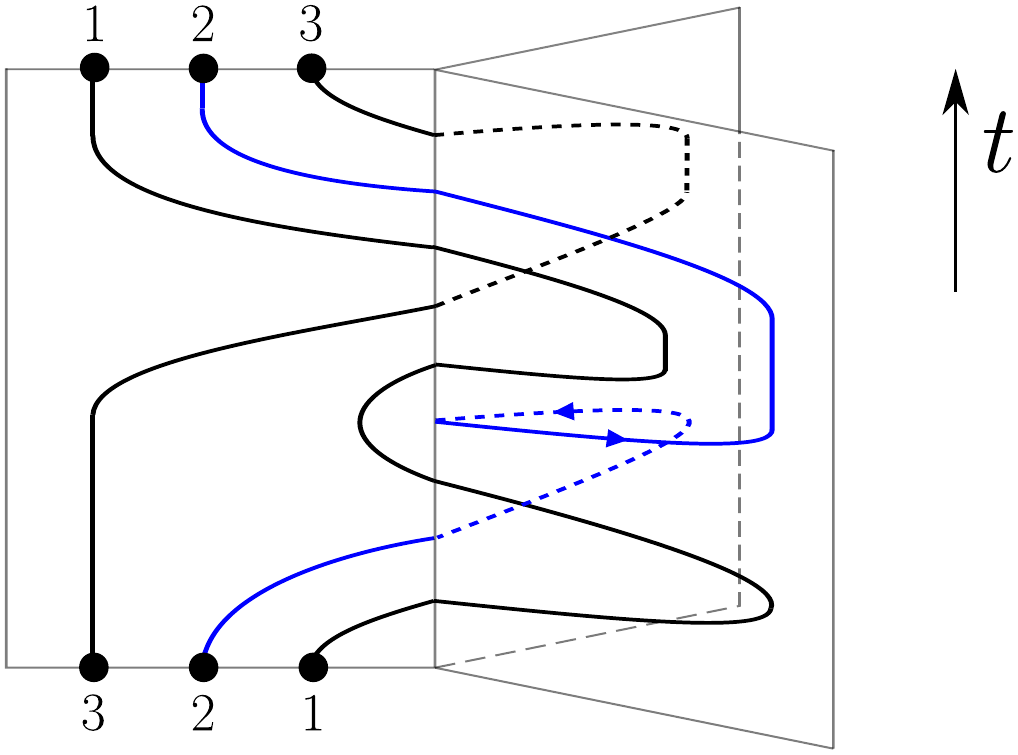}
\caption{Braid $\sigma_1^{(2,1)}\sigma_2^{(2,2,1)}\sigma_1^{(2,1)}$ on a trijunction. The depicted braid has been deformed to simplify the picture.}
\label{norel}
\end{figure}

In order to visualise the counterparts of braids in the braid relation from Fig.\ref{braid}, we need to define the counterpart of $\sigma_2$ - the simple braid exchanging anyons $2$ and $3$. To this end, anyon $1$ has to be moved to the right or left branch of the junction so that anyons  $2$ and $3$ can carry on and exchange as on Fig. \ref{comparison}b. Let us choose the braid where anyon $1$ moves to the right branch, which we denote $\sigma_2^{(2,2,1)}$. Composition $\sigma_1^{(2,1)}\sigma_2^{(2,2,1)}\sigma_1^{(2,1)}$ is shown on Fig. \ref{norel}.
Strikingly, the world line of the middle anyon (blue line on Fig. \ref{norel}) is now blocked by world lines of the other particles and it cannot be deformed freely. Consequently, there is no braiding relation on a trijunction. In fact, the three-particle braid group of the trijunction is freely generated by $\sigma_1^{(2,1)}$, $\sigma_2^{(2,2,1)}$ and $\sigma_2^{(1,2,1)}$. However, when one considers a bigger junction or a larger number of anyons, some relations appear. Their precise form is as follows.
\begin{enumerate}
\item For $n\geq 4$, pseudo-commutative relations appear. For $j-i\ge 2$,
\begin{equation}\label{equation:pseudo-commutative}
\sigma_j^{a_1\dots a_{j+1}}\sigma_i^{a_1\dots a_{i+1}} = \sigma_i^{a_1\dots a_{i+1}}\sigma_j^{a_1\dots a_{i-1}a_{i+1}a_ia_{i+2}\dots a_{j+1}}.
\end{equation}
\item For $d\geq 4$ and $n\geq 3$, pseudo-braid relations appear. For $1\le i\le n-2$,
\begin{align}\label{equation:pseudo-braid}
&\mathrel{\hphantom{=}}\sigma_{i+1}^{a_1\dots a_{i-1}a_i a_{i+1}a_{i+2}}\sigma_i^{a_1\dots a_{i-1}a_i a_{i+2}}\sigma_{i+1}^{a_1\dots a_{i-1}a_{i+2}a_i a_{i+1}}\notag\\
&=\sigma_i^{a_1\dots a_{i-1}a_i a_{i+1}}\sigma_{i+1}^{a_1\dots a_{i-1}a_{i+1}a_i a_{i+2}}\sigma_i^{a_1\dots a_{i-1}a_{i+1}a_{i+2}}.
\end{align}
\end{enumerate}

Let us emphasise that the braid group of a $d$-junction has more generators than the planar braid group and hence it imposes fewer topological constraints on the unitary braiding operators that are assigned to simple braids in a physical model. This is perhaps most striking in the case of three anyons on a trijunction where we had three generators and no relations between them.

{\it General (planar) network architectures.} -- In order to relate braiding relations for anyons on general networks with the braiding of anyons in 2D, we first have to consider a different presentation of the planar braid group. Namely, we will consider the total braid, $\delta$, which is a product of all simple braids, $\delta:=\sigma_{1}\sigma_{2}\hdots\sigma_{n-1}$. Braid $\delta$ corresponds to the move where the first anyon exchanges consecutively with all anyons. Using 2D braiding relations one can show that any simple braid can be expressed by $\sigma_1$ and $\delta$ as $\sigma_i=\delta^{i-1}\sigma_1\delta^{1-i}$ \cite{braids}. 

\begin{figure}
\centering
\includegraphics[width=0.45\textwidth]{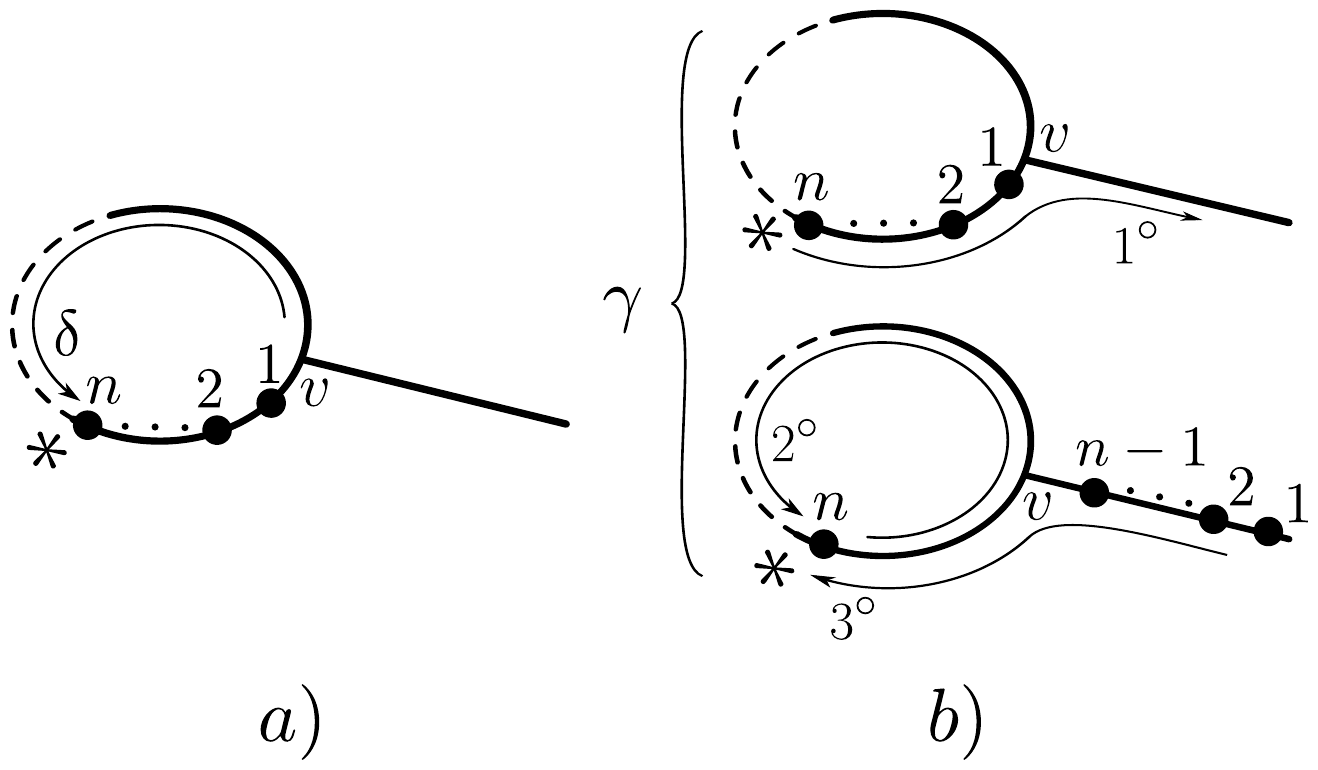}
\caption{a) Total braid $\delta$ on a lollipop network. b) One-particle move $\gamma$ on a lollipop network. The rooted spanning tree with root $*$ is drawn by solid lines.}
\label{lollipop}
\end{figure}

We will next show how the above relations are recovered on networks. To this end, we fix a spanning tree of our network, $T$, which is a connected tree that contains all vertices of the network. Moreover, we choose the root of $T$ to be a vertex of degree two that lies on the boundary of the network. The initial configuration of anyons is such that the anyons are assembled on the edge of $T$ which is incident to the root. The above choice of a spanning tree unambiguously defines all possible exchanges on junctions. To see this, note that for every essential vertex $v$ of the network (i.e. vertex at which three or more edges are incident), we have a unique path in $T$, denoted by $[v,*]$, that connects this vertex with the root. Such a path implies labelling of branches of the junction at $v$ with branch $0$ being the one that is contained in $[v,*]$ and the remaining branches labelled clockwise as described in the previous section. Consequently, simple braids at $v$ will be denoted by additional superscript -- $v$. The counterpart of total braid $\delta$ is realised by utilising a loop containing the root of $T$ (effectively considering a lollipop-shaped subnetwork, see Fig.\ref{lollipop}) -- anyon $1$ is transported along the loop to the end of the line.
It is straightforward to check that up to some backtracking moves, in the lollipop setting from Fig. \ref{lollipop} we have 
\begin{equation}\label{eq:lollipop}
\sigma_i^{v;(1,\hdots,1,2,1)}=\delta^{i-1}\sigma_1^{v;(2,1)}\delta^{1-i}.
\end{equation}
Moreover, braid $\delta$ can be expressed in terms of simple braids at the junction of the lollipop and a one-particle move $\gamma$ defined as the move where anyons $1$ through $n-1$ are transported to branch $2$ of the junction and anyon $n$ travels alone around the lollipop loop. The precise relation reads 
\begin{equation}\label{eq:delta-prod}
\gamma=\sigma_{n-1}^{v;(2,\hdots,2,2,1)}\hdots\sigma_1^{v;(2,1)}\delta. 
\end{equation}

Let us pause for a moment to analyse the role of one-particle moves. Such moves do not describe any exchange, hence assigning unitary operators to these moves can only come from the existence of some external gauge fields puncturing the plane where the considered network is confined. For instance, the presence of a delta-like magnetic flux flowing perpendicularly through the middle of the lollipop loop would result with multiplication of the anyonic wave function by a phase factor due to the Aharonov-Bohm effect. From now on, we will always assume that there are no such external gauge fields present in the system. Consequently, we will equate all one-particle loops to identities. 

By putting $\gamma$ to identity, we obtain $\delta=\sigma_1^{v;(1,2)}\hdots\sigma_{n-1}^{v;(2,\hdots,2,1,2)}$. Note that at this point we have almost recovered the presentation of the planar braid group that we considered at the beginning of this section. The only difference is that expression for $\delta$ involves different simple braids than expression (\ref{eq:lollipop}). As we show in the next section, this problem disappears for a wide class of networks that are sufficiently connected. 

We say that a network is $k$-connected when any two of its essential vertices can be connected by at least $k$ paths that are mutually internally disjoint. By Menger's theorem \cite{menger}, this is equivalent to the fact that after removing at most $k-1$ vertices, the network remains connected.

{\it Braiding on $2$-connected networks.} -- The key feature of $2$-connected networks that simplifies their braid groups is that for every trijunction in the network we can find suitable lollipop subnetworks that allow us to reduce the number of generators. In particular, if $v$ is the first essential vertex from the root $*$ (i.e. there are no essential vertices on path $[v,*]$, as in Fig.\ref{lollipop}), then for every branch $a_1$ at junction $v$, there exists a path connecting $v$ and $*$ that contains branch $a_1$ and is independent of $[v,*]$. Consequently, we have a lollipop where, for any $\ba=(a_1,a_2,\hdots,a_{i+1})$, we obtain
\begin{equation}\label{eq:simple-delta}
\sigma_i^{v;\ba}=\delta\sigma_{i-1}^{v;\ba'}\delta^{-1},
\end{equation}
where $\ba'=(a_2,\hdots,a_{i+1})$. The above expression allows us to inductively reduce any simple braid at $v$ to a braid of the form $\delta^{i-1}\sigma_{1}^{v;(a,b)}\delta^{1-i}$ with $a>b$. This in turn means that for simple braids taking place at a fixed trijunction spanned on branches $(a,b)$ at vertex $v$, we indeed obtain a set of 2D braiding relations. However, braids at different trijunctions are still {\it a priori} independent of each other. One can show that a similar situation concerns simple braids at junctions that are further away from the root.

\begin{figure}[h]
\centering
\includegraphics[width=0.45\textwidth]{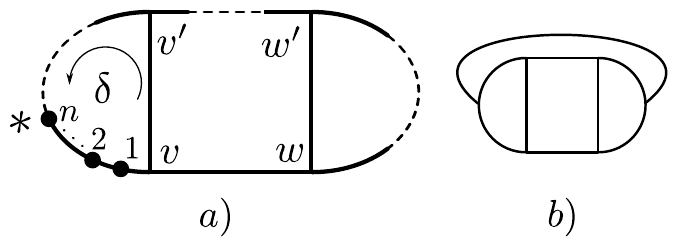}
\caption{a) A modular network with two modules containing junctions $v, v'$ and $w,w'$ respectively. b) Network from point a) modified by adding one edge that makes it $3$-connected. }
\label{modular}
\end{figure}

To recapitulate, $2$-connected networks indeed support genuine 2D braiding relations. However, the relations are valid only within certain sets of braids that are restricted to fixed trijunctions of the network. This is strikingly different from the 2D anyon braiding where braiding is ruled by only one type of simple braids. Note that this feature of braiding can be utilised to design networks consisting of different modules where quantum statistics can be changed when moving anyons from module to module. An example of such a modular network is shown on Fig. \ref{modular}.

\begin{figure}
\centering
\includegraphics[width=0.35\textwidth]{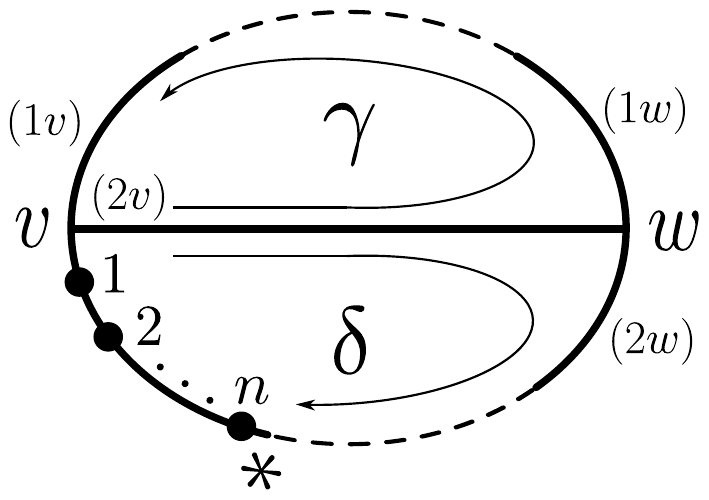}
\caption{A $\Theta$-network with the rooted spanning tree marked by solid lines. Arrows symbolise the total braid $\delta$ and one-particle loop $\gamma$ as described in the main text.}
\label{theta}
\end{figure}

{\it Braiding on $3$-connected networks.} -- In contrast to the great complexity of braiding scenarios outlined so far for $1$- and $2$-connected networks, $3$-connected networks bring a tremendous simplification. The key is to consider the so-called $\Theta$-network that consists of two essential vertices connected by three edges (Fig. \ref{theta}). By definition, every two essential vertices in a $3$-connected network can be connected by three independent paths that form a $\Theta$-subgraph. The key property of anyon braiding on a $\Theta$-network is that it identifies simple braids on different trijunctions. This in turn means that the 2D braiding relations are recovered. Similarly, by considering suitable $\Theta$-subnetworks of a general $3$-connected network, one can show that the above identification recovers the planar braid group. Let us see explicitly how it happens for a $\Theta$-network from Fig.\ref{theta}.
Denote by $\gamma$ the one-particle loop where the first anyon travels around the upper loop on Fig. \ref{theta}. Moreover, denote by $\gamma'$ a move which involves anyons $1$ and $2$ where i) anyon $1$ travels to branch $(2w)$ through the solid edge $[v,w]$, ii) anyon $2$ travels around the upper loop and iii) anyon $1$ goes back along the tree from branch $(2w)$. Up to some backtracking moves, we have the following relations.
\begin{equation}\label{theta-rel}
\delta\gamma=\gamma' \delta,\quad \sigma_1^{w;(2,1)}\gamma=\gamma' \sigma_1^{v; (2,1)}.
\end{equation}
Because $\gamma$ is a one-particle move, according to our assumption about the non-existence of external fields, we put it to identity. Then, the left relation in (\ref{theta-rel}) implies that $\gamma'$ is identity as well. This in turn applied to the right relation yields $\sigma_1^{w;(2,1)}=\sigma_1^{v; (2,1)}$.

To sum up, relations (\ref{eq:delta-prod}) and (\ref{eq:simple-delta}) for lollipops together with relation (\ref{theta-rel}) for $\Theta$-subnetworks enabled us to identify the {\it a priori} complicated braiding relations on networks with the well-known 2D braiding when the considered network is $3$-connected. 

{\it Example: a modular $2$-connected network.} -- A simple realisation of a modular network that admits different non-abelian quantum statistics in different modules is shown on Fig. \ref{modular}. 
Because one can span a $\Theta$-connection between trijunctions at $v$ and $v'$, simple braids at these junctions are identified with each other. Similarly, one identifies simple braids at trijunctions at $w$ and $w'$. However, simple braids at $v$ and $w$ are independent of each other. Moreover, appropriate lollipop relations ensure that $\sigma_i^{v;\ba}=\delta^{i-1}\sigma_1^{v;(2,1)}\delta^{1-i}$ and $\sigma_i^{w;\ba}=\delta^{i-1}\sigma_1^{w;(2,1)}\delta^{1-i}$ for any $\ba$, hence braiding is independent of the distribution of anyons. There are no topological constraints that would forbid utilising simple braids at $v$ and $w$ in such a way that they would realise different topological quantum gates $\sigma_i^{v;\ba}\to U_i$, $\sigma_i^{w;\ba}\to V_i$. In such a system, braiding of anyons $i$ and $(i+1)$ at junctions $v$ or $v'$ would realise gate $U_i$, while braiding at junctions $w$ or $w'$ would realise gate $V_i$. This proposal shows that conducting quantum computations on a topological quantum computer based on such a modular network would be a relatively easy task provided that one could manipulate anyons efficiently. 

Finally, let us remark that the above desired features of the modular network are lost when one adds just a single edge to the network in the way shown on Fig. \ref{modular}b. By a visual inspection, one can check that network from Fig.\ref{modular}b is now $3$-connected. Analogous properties of quantum statistics on networks have been observed for abelian anyons \cite{HKRS,KoPark}.

{\it Relation to the braiding of Majorana fermions} -- Let us consider Majorana fermions in quantum wires modelled by Kitaev superconducting chain of spinless fermions \cite{kitayev,oppen-notes}. In a certain range of this model's hamiltonian (called the topological region forming topological strings) there exists one zero-energy eigenmode $d_0$ which can be represented in terms of two Majorana edge modes $\gamma_1$ and $\gamma_2$, $d_0=\gamma_1+i\gamma_2$, that are localised at the beginning and at the end of the chain respectively \cite{kitayev}. Because of the localisation of Majorana modes, one can consider their braiding on chains that are coupled into trijunctions by adiabatically tuning parameters of the hamiltonian in a local fashion \cite{alicea}. Performing quantum computations with Majorana edge modes would require creating a network with multiple well-separated edge modes on it. It has been shown in \cite{alicea} that exchange of two edge modes $\gamma_{i}$ and $\gamma_{i+1}$ gives quantum gate $U_i=\exp(\pi \gamma_i\gamma_{i+1}/4)$. Moreover any one-particle move where just one Majorana fermion is being adiabatically transported, results with the multiplication of the wavefunction by a global phase factor. Hence, in terms of anyon braiding, this model has the following properties i) all one-particle moves do not change the quantum state of the system (are effectively put to identity) and ii) all simple braids are represented by the same quantum gate, i.e. $\sigma_i^{v;\ba}\to U_i$ for any $v$ and $\ba$. Although our introduced braids ignore the existence of topological regions that connect pairs of Majorana edge modes, our approach can be adapted to take them into account. In particular, one can perform all braids on junctions, all lollipop moves and $\Theta$-moves in a way that avoids self-intersections of the toplogical 'strings' (see Appendix). Therefore, braiding of Majorana fermions on any network is exactly the same as braiding in 2D and it seems not to exploit the full potential of modular networks outlined in previous sections. That said, this model shows that one can hope to find other physical models for anyon braiding on networks that would not be directly equivalent do 2D braiding.

\begin{acknowledgments}
The authors gratefully acknowledge the support of the American Institute of Mathematics (AIM) where this collaboration was initiated. We would like to thank Adam Sawicki and Jon Harrison for useful discussions during the workshop at AIM. TM wold like to also thank Nick Jones and Joost Slingerland for discussions about Majorana fermions and physical aspects of anyon braiding and Jonathan Robbins for helpful feedback concerning the manuscript. Byung Hee An was supported by the National Research Foundation of Korea (NRF) grant funded by the Korea government (MSIT) No. 2020R1A2C1A01003201. TM acknowledges the support of the Foundation for Polish Science (FNP), START programme.
\end{acknowledgments}

\onecolumngrid
\appendix*
\section{Appendix -- Braiding of Majorana Fermions on Quantum Wire Networks}

Let us consider a particular model that supports braiding of anyons on networks. It describes a network of spinless fermionic quantum chains that are coupled together at their endpoints forming a network of trijunctions. Hamiltonian of a single quantum chain is given by 
\begin{equation}\label{ham}
H=-\mu\sum_{k=1}^N c_k^\dagger c_k-\sum_{k=1}^{N-1}\left(tc_k^\dagger c_{k+1}+|\Delta|e^{i\phi}c_k c_{k+1}+h.c.\right)
\end{equation}
where parameters $\mu$ and $t>0$ are respectively the chemical potential and hopping amplitude $t$. Parameter $|\Delta|e^{i\phi}$ is called the superconducting gap, lending itself to the role of the model in the description of spinless $p$-wave superconductors \cite{oppen-notes}.  When $|\mu|<2t$, hamiltonian (\ref{ham}) for $N\to\infty$ has one zero-energy eigenmode $d_0$ which can be represented in terms of two Majorana modes $\gamma_1$ and $\gamma_2$, $d_0=\gamma_1+i\gamma_2$, that are localised at the beginning and at the end of the chain respectively \cite{kitayev}. This property persists for finite chains with mode $d_0$ having energy which is exponentially small in the size of the chain. As Majorana modes $\gamma_1$ and $\gamma_2$ are localised at the endpoints of the chain, one can consider their braiding on chains that are connected into trijunctions by adiabatically tuning parameters of hamiltonian (\ref{ham}) in a local fashion \cite{alicea}. As authors of \cite{alicea} point out, braiding is well-defined provided that i) $\gamma_1$ and $\gamma_2$ remain sufficiently separated throughout the process and ii) the energy gap remains open at all times. Schematically, one can visualise adiabatic transport of Majorana fermions as a transport of two point-like anyons that are connected by a string of topological region where parameters of the hamiltonian satisfy $|\mu|<2t$. Performing quantum computations with Majorana edge modes would require creating a network with multiple well-separated topological regions on it. Creating $n$ topological regions results with $2n$ edge modes which can be subsequently braided on junctions. It has been shown in \cite{alicea} that exchange of two edge modes $\gamma_{i}$ and $\gamma_{i+1}$ gives quantum gate $U_i=\exp(\pi \gamma_i\gamma_{i+1}/4)$. Moreover any one-particle move where just one Majorana fermion is being adiabatically transported, results with the multiplication of the wavefunction by a global phase factor. Hence, in terms of anyon braiding, this model has the following properties i) all one-particle moves do not change the quantum state of the system (are effectively put to identity) and ii) all simple braids are represented by the same quantum gate, i.e. $\sigma_i^{v;\ba}\to U_i$ for any $v$ and $\ba$. 

Let us next briefly review how simple braids are realised with such Majorana fermions, as described in \cite{alicea}. Recall that a simple braid is described by $\sigma_i^{v;\ba}$ where $\ba=(a_1,\hdots,a_{i+1})$ denotes branches of junction $v$ that accommodate anyons that are moved during the exchange (we always have $a_k\in \{1,2\}$ because the Majorana model so far has been developed for trijunctions only). In particular, $a_1,\hdots,a_{i-1}$ denote branches where the top $(i-1)$ anyons are distributed before anyons $i$ and $i+1$ exchange on branches $a_i$ and $a_{i+1}$.  First of all, we want to avoid such distributions of the top $(i-1)$ anyons where a topological string is stretched between different branches. In other words, we consider only such $\ba$ where $a_1=a_2$, $a_3=a_4$, $\hdots$, $a_{i-1}=a_{i-2}$ for $i$ odd and $a_{i-2}=a_{i-3}$ for $i$ even. This means that we are in fact considering a subgroup of the full graph braid group generated by the above specific generators. This, however, is not a big constraint. In particular, in the case of $n=3$ anyons on a trijunction (understood as the situation where the fourth Majorana fermion is always kept at the fixed position next to the root), we do not throw away any generators and the modified braid group is the same as the standard graph braid group, i.e. it is freely generated by $\sigma_1^{(2,1)}$, $\sigma_2^{(1,2,1)}$ and $\sigma_2^{(2,2,1)}$. 

\begin{figure}[h!]
\centering
\includegraphics[width=\textwidth]{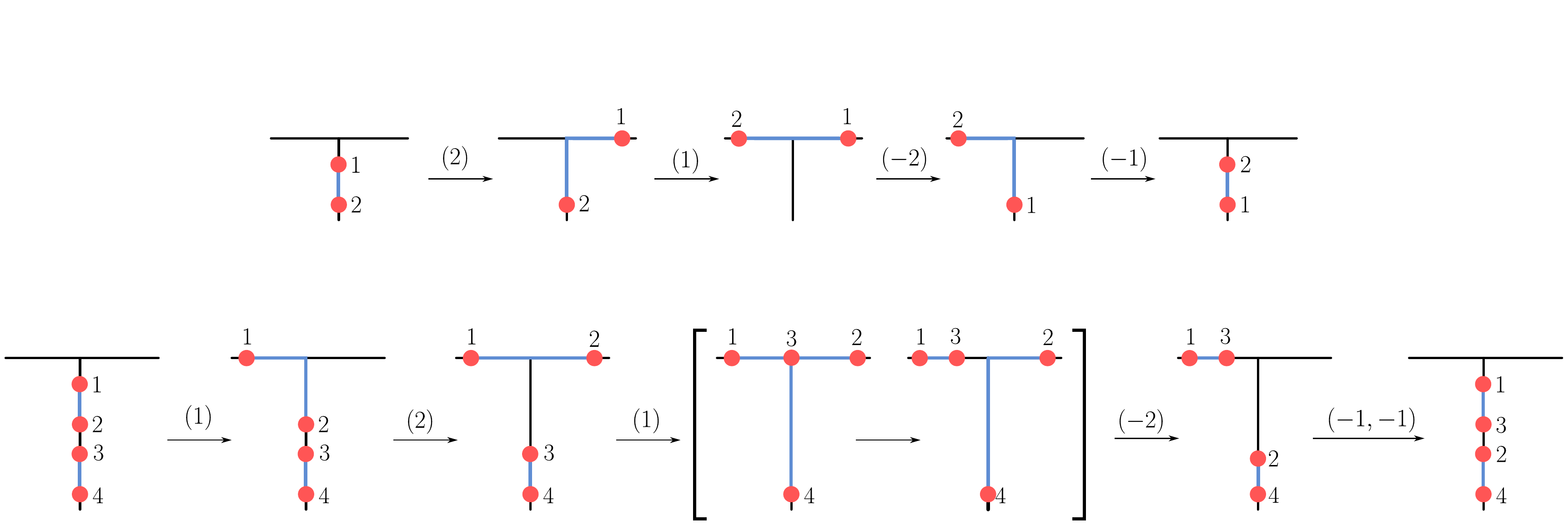}
\caption{Simple braids realised by Majorana fermions depicted as sequences of 'stop-frames'. Topological strings are depicted by blue lines. The top row shows exchange $\sigma_1^{(2,1)}$. The left branch of the junction is denoted by $(1)$ and the right branch is denoted by $(2)$. Numbers over arrows between frames denote the move that has been done. For instance, $(2)$ denotes the move where the top anyon travels to the right branch and $(-2)$ is the inverse move. The bottom row shows exchange $\sigma_2^{(1,2,1)}$ during which two topological strings intersect (in square brackets). Compare with Fig.~3 in \cite{alicea}.}
\label{simple-m}
\end{figure}

The above described simple braids $\sigma_i^{v;\ba}$ with Majorana fermions are realised in different ways, depending on whether $i$ is even or odd. Simple braids $\sigma_i^{v;\ba}$ with $i$ odd correspond to situations when we want to perform an exchange of two anyons connected by a topological string. The second type of braids concerns $i$ even and corresponds to situations when Majorana fermions from two different topological regions are exchanged and strings cross each other during the exchange process. Let us fix the orientation of simple braids so that $a_i=2$ and $a_{i+1}=1$ with our standard notation in which $2$ denotes the right branch and $1$ the left branch. Examples of the two types of exchanges are shown in Fig.~\ref{simple-m}.
It is fairly straightforward to see that for the above truncated set of simple braids on a trijunction the pseudo-commutative relations become ordinary commutative relations.

The next difficulty is to rewrite different lollipop and $\Theta$-relations so that they make sense for Majorana fermions that are pairwise connected by topological strings. In particular, we want to 
\begin{itemize}
\item avoid configurations of anyons where a single topological string intersects itself and
\item consider moves where the initial configuration of topological strings is the same as the final configuration of topological strings.
\end{itemize}
For instance, such situations appear when one Majorana fermion travels around a simple loop of the network. The key idea is to rewrite lollipop and $\Theta$-relations in such a way that they are homotopy equivalent to relations consisting of moves where topological strings do not self-intersect anymore. In the remaining part of this Supplemental Material we show that this is (almost) always possible, except for the $\Theta$-relation which has to be modified a bit more.

\begin{figure}[h!]
\centering
\includegraphics[width=\textwidth]{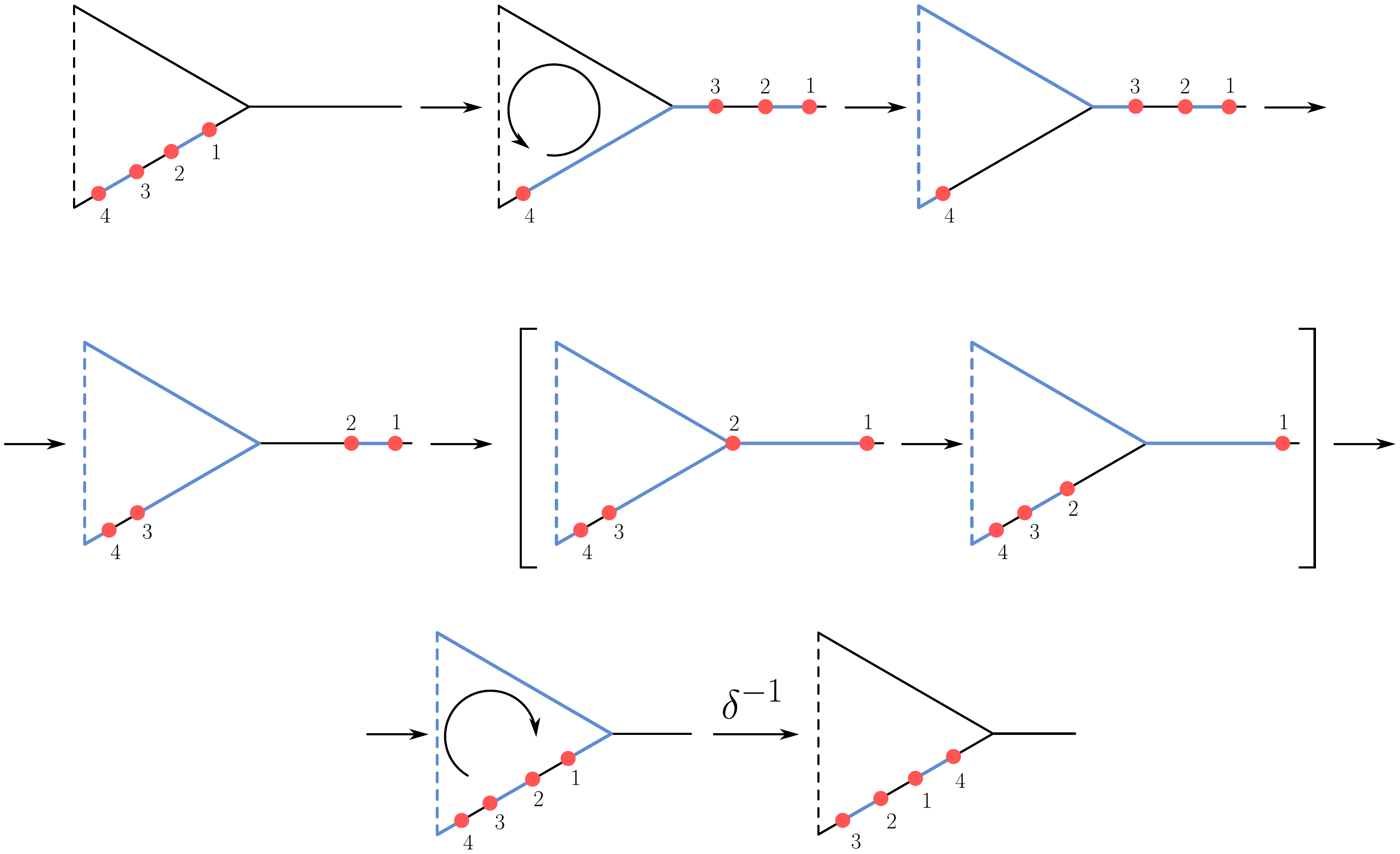}
\caption{Lollipop move $\gamma\delta^{-1}$ for $n=4$ Majorana fermions. Importantly, the move involves a crossing of two topological strings (square brackets). Topological strings are depicted by blue lines. }
\label{gammadelta}
\end{figure}

Let us start with the lollipop relation which for 'free' anyons reads $\gamma=\sigma_{n-1}^{v;(2,\hdots,2,2,1)}\hdots\sigma_1^{v;(2,1)}\delta$. By a visual inspection of the above moves, one can see that move $\gamma$ changes the configuration of topological strings. This can be easily remedied by rewriting the above relation as
\begin{equation}\label{lollipop1-m}
\gamma\delta^{-1}=\sigma_{n-1}^{v;(2,\hdots,2,2,1)}\hdots\sigma_1^{v;(2,1)}.
\end{equation}
The RHS of \eqref{lollipop1-m} clearly satisfies our conditions. To convince ourselves that the LHS of \eqref{lollipop1-m} does it too, let us inspect move $\gamma\delta^{-1}$ for $n=4$ on Fig.~\ref{gammadelta}.

\begin{figure}[h!]
\centering
\includegraphics[width=\textwidth]{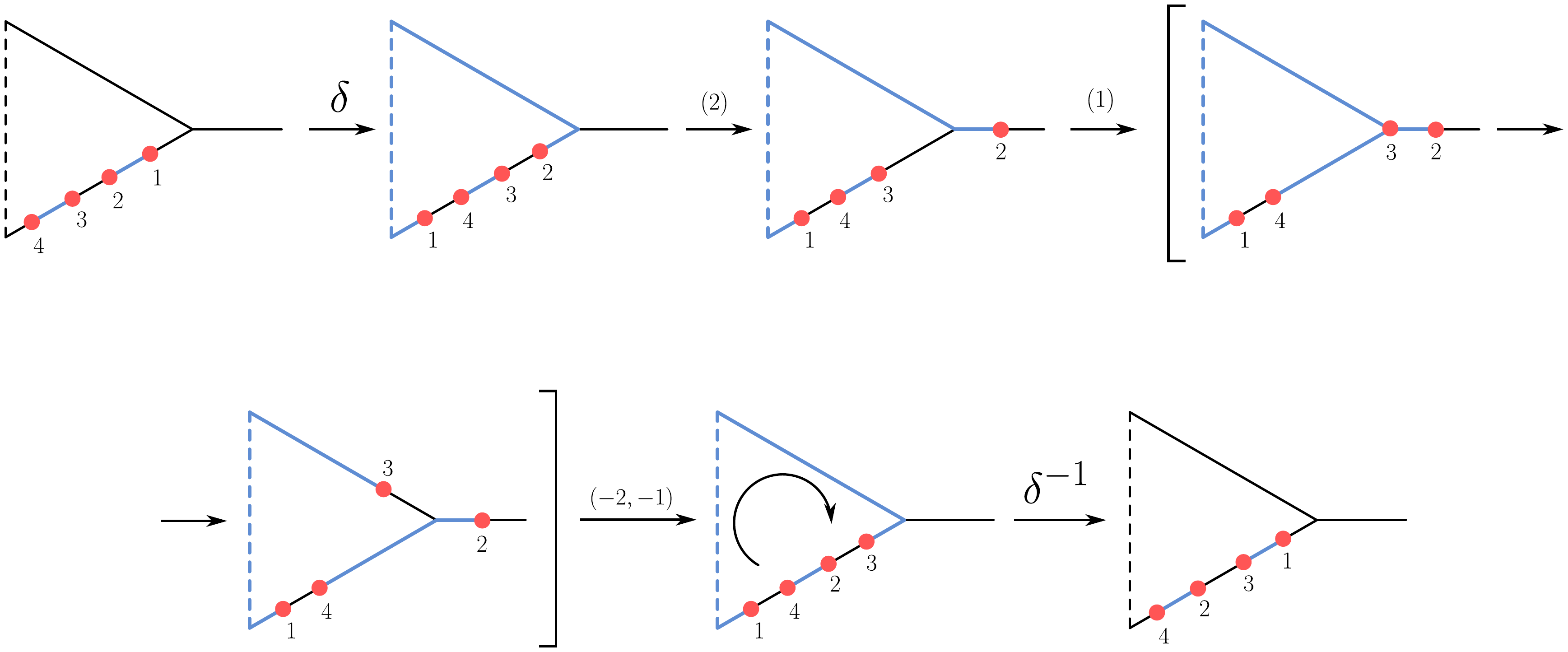}
\caption{Lollipop braid $\delta\sigma_1^{v;(2,1)}\delta^{-1}$. Square brackets mark the point of intersection of two topological strings. Topological strings are depicted by blue lines.}
\label{lollipop2}
\end{figure}

Similarly, one can check that the RHS of lollipop relation
\begin{equation}\label{lollipop2-m}
\sigma_i^{v;(1,\hdots,1,2,1)}=\delta^{i-1}\sigma_1^{v;(2,1)}\delta^{1-i}
\end{equation}
indeed preserves the configuration of topological strings and avoids self-intersections of strings. This is obvious for $i$ odd as $\delta^2$ itself preserves the configuration of topological regions. For $i$ even, it is helpful to inspect relation $\sigma_2^{v;(1,2,1)}=\delta\sigma_1^{v;(2,1)}\delta^{-1}$ in Fig.~\ref{lollipop2}.

The next step is to deduce braiding relations from the above lollipop relations \eqref{lollipop1-m} and \eqref{lollipop2-m}. For 'free' anyons this has been done simply by putting one-particle loops to identities. However, for Majorana fermions we want to avoid the problematic one-particle loops. This is done directly by combining \eqref{lollipop1-m} with \eqref{lollipop2-m} to obtain
\begin{equation}\label{almost-braid}
\sigma_i^{v;(1,\hdots,1,2,1)}=\left(\sigma_1^{v;(1,2)}\hdots\sigma_{n-1}^{v;(2,\hdots,2,1,2)}\gamma\right)^{i-1}\sigma_1^{v;(2,1)}\left(\gamma^{-1}\sigma_{n-1}^{v;(2,\hdots,2,2,1)}\hdots\sigma_1^{v;(2,1)}\right)^{i-1}.
\end{equation}
Our claim is that modulo one-particle moves that do not involve self-intersections of topological strings, RHS of equation \eqref{almost-braid} preserves the configuration of topological strings and does not contain self-intersections of strings and is equivalent to $\Delta_n^{1-i}\sigma_1^{v;(2,1)}\Delta_n^{i-1}$ with $\Delta_n:=\sigma_{n-1}^{v;(2,\hdots,2,2,1)}\hdots\sigma_1^{v;(2,1)}$. Our proof of the above claim is inductive with respect to $i$. First, we rewrite the RHS of \eqref{almost-braid} as
\[\Delta_n^{-1}\gamma\hdots\Delta_n^{-1}\gamma\Delta_n^{-1}\left(\gamma\sigma_1^{v;(2,1)}\gamma^{-1}\right)\Delta_n\gamma^{-1}\Delta_n\hdots\gamma^{-1}\Delta_n.\]
Next, we note that move $\gamma\sigma_1^{v;(2,1)}\gamma^{-1}$ i) preserves the configuration of topological strings and does not contain self-intersections of strings, ii) up to one-particle moves is equivalent to $\sigma_1^{v;(2,1)}$. This in particular means that relation \eqref{almost-braid} for $i=2$ is well-defined for Majorana fermions and yields $\sigma_2^{v;(1,2,1)}\sim \Delta_n^{-1}\sigma_1^{v;(2,1)}\Delta_n$ modulo one-particle moves. Hence, we get that the RHS of \eqref{almost-braid}, up to single-particle moves, is equivalent to 
\[\Delta_n^{-1}\gamma\hdots\Delta_n^{-1}\gamma\Delta_n^{-1}\left(\gamma\sigma_2^{v;(1,2,1)}\gamma^{-1}\right)\Delta_n\gamma^{-1}\Delta_n\hdots\gamma^{-1}\Delta_n\]
with left and right strings of operators appropriately shorter. Next, we repeat our reasoning for move $\gamma\sigma_2^{v;(1,2,1)}\gamma^{-1}$ and continue up to $i=n-1$. This ends the proof of our claim. Finally, to prove the existence of true braiding relations for Majorana fermions with topological strings on $2$-connected networks, we repeat the reasoning from the main text noting that $\gamma\delta^{-1}$ associated with different simple cycles of the network differ only by well-defined one-particle moves. Hence, expressions $\Delta_n$ associated with different distributions of anyons on a fixed trijunction of a $2$-connected network yield the same braiding operations.

\begin{figure}[h!]
\centering
\includegraphics[width=.5\textwidth]{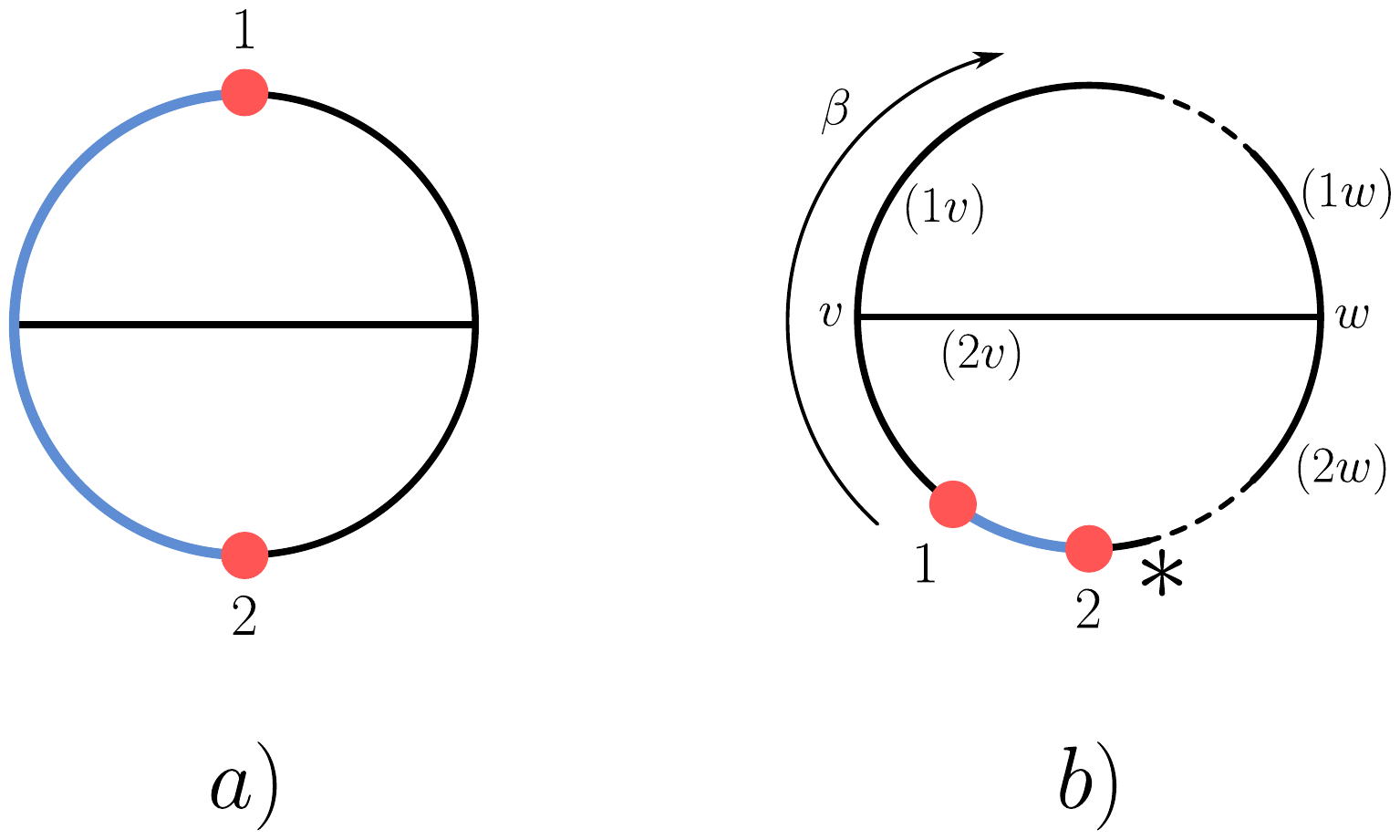}
\caption{a) The symmetric initial configuration of Majorana fermions for the $\Theta$-relation. b) The underlying rooted spanning tree (solid lines) with root $*$ that defines simple braids $\sigma_1^{v;2,1}$ and $\sigma_1^{w;2,1}$ as described in the main text. Move $\beta$ transforms the standard initial configuration to the symmetric configuration in a). Topological strings are depicted by blue lines.}
\label{theta-init}
\end{figure}

\begin{figure}[h!]
\centering
\includegraphics[width=.8\textwidth]{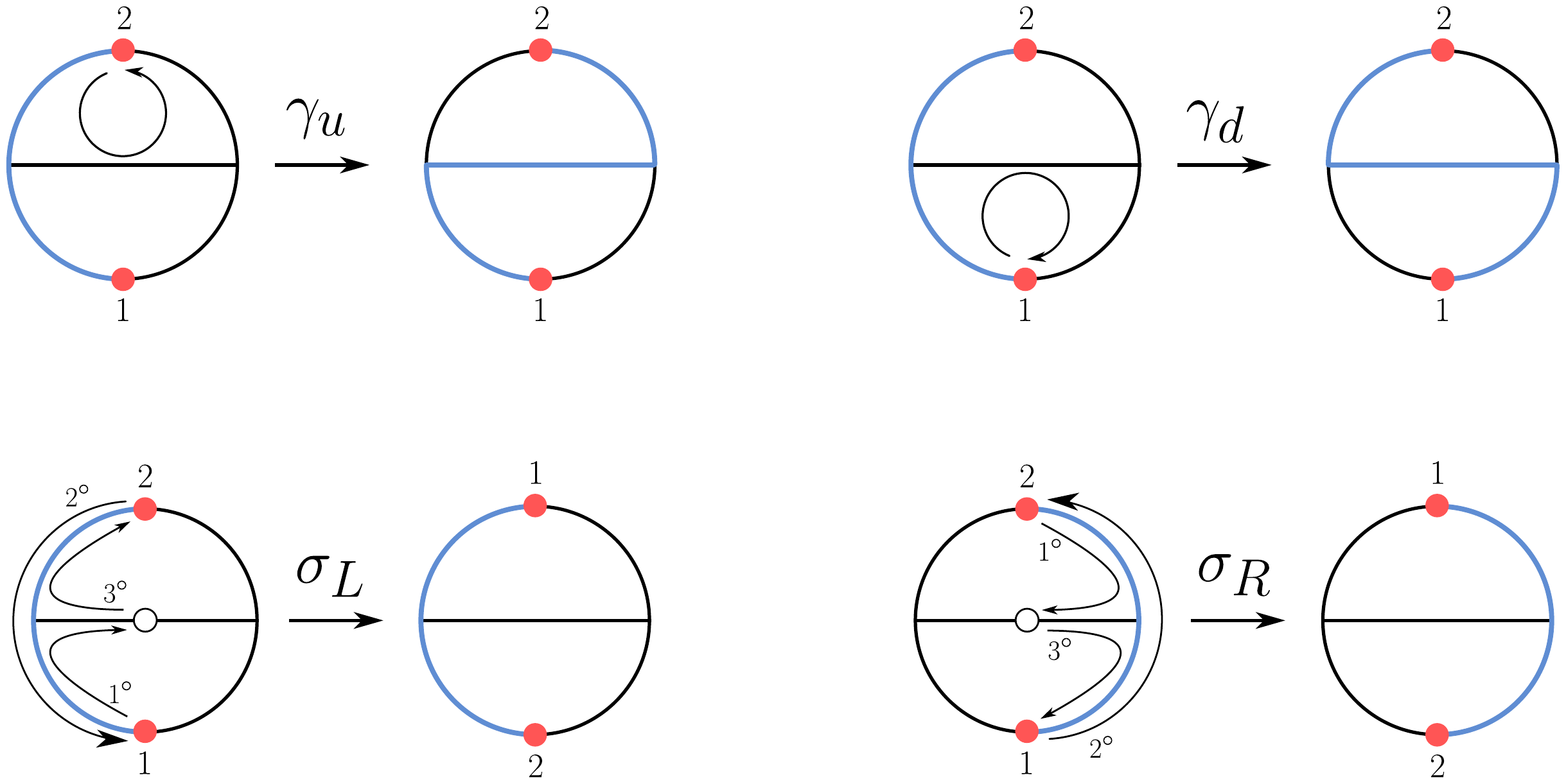}
\caption{Pictorial definitions of single-particle moves $\gamma_u$ and $\gamma_d$ and exchanges $\sigma_L$ and $\sigma_R$ that will be used to derived the $\Theta$-relation for Majorana fermions. Topological strings are depicted by blue lines.}
\label{theta-moves}
\end{figure}

\begin{figure}[h!]
\centering
\includegraphics[width=.6\textwidth]{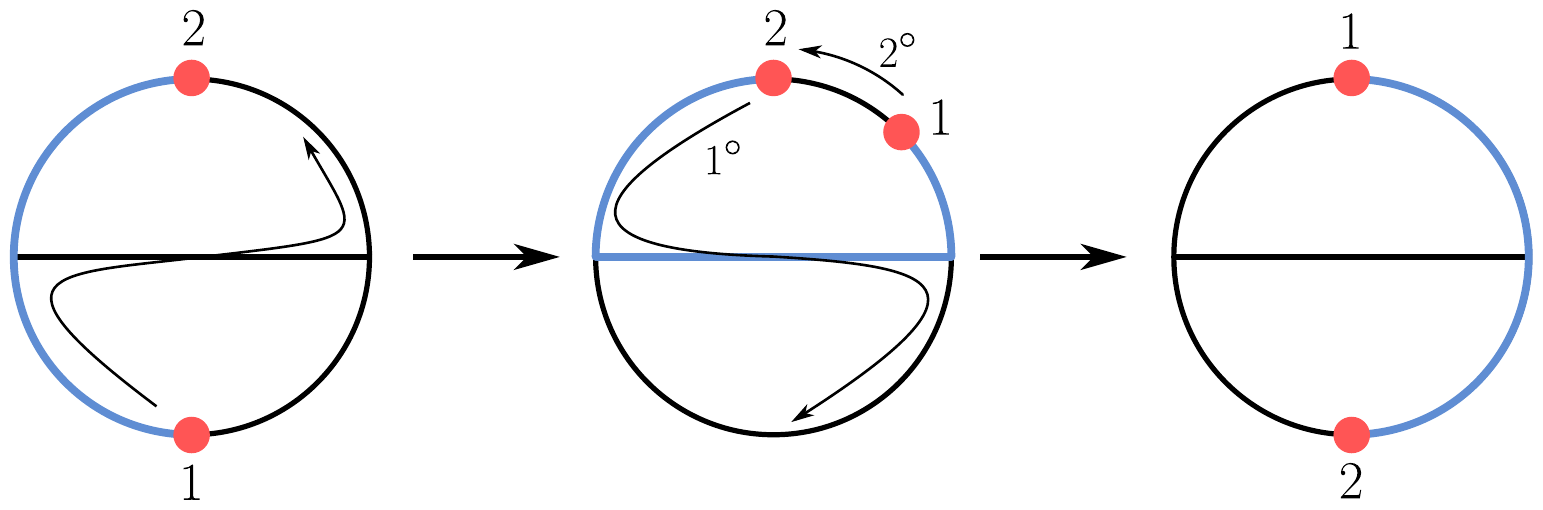}
\caption{A move which is homotopy equivalent to both $\sigma_L\gamma_u\gamma_d$ and $\gamma_d\gamma_u\sigma_R$. Topological strings are depicted by blue lines.}
\label{homotopy}
\end{figure}

As the last part of this Supplemental Material, let us adapt the $\Theta$-relation to the situation of Majorana fermions connected by topological strings. Our goal is to argue that in the case of $3$-connected networks exchanges on different junctions of the network give the same braiding operators for Majorana fermions with strings. Let us start with just one topological string, i.e. $n=2$. In contrast to our standard setting, the starting position for all moves will be such that anyon $1$ will be located at the top part of the $\Theta$-network and anyon $2$ will be located symmetrically at the bottom of the network, see Fig.~\ref{theta-init}a). Such an initial configuration can be easily obtained from the standard initial configuration by performing a single-particle move $\beta$ shown on Fig.~\ref{theta-init}b).
Moreover, we define single-particle moves $\gamma_u$ and $\gamma_d$ as well as exchanges $\sigma_L$ and $\sigma_R$ as shown on Fig.~\ref{theta-moves}.
Next, we consider the following two composite moves: $\sigma_L\gamma_u\gamma_d$ and $\gamma_d\gamma_u\sigma_R$. The key step is to note that both of the composite moves are homotopy equivalent,
\begin{equation}\label{theta-homotopy}
\sigma_L\gamma_u\gamma_d\cong\gamma_d\gamma_u\sigma_R.
\end{equation}
In other words, there exists a continuous deformation of moves between the two composite moves in \eqref{theta-homotopy}. Moreover, the deformation is such that it avoids self-crossings of the topological string. In particular, both composite moves can be continuously deformed to the move shown in Fig.~\ref{homotopy}.
Hence, the composite move $\beta\sigma_L\gamma_u\gamma_d\left(\gamma_d\gamma_u\sigma_R\right)^{-1}\beta^{-1}$ is homotopy equivalent to a trivial braid that preserves the configuration of the topological string. Moreover, up to single-particle moves, we have that
\[\beta\sigma_L\beta^{-1}\sim\sigma_1^{v;1,2},\quad \beta\gamma_u\gamma_d\left(\gamma_d\gamma_u\sigma_R\right)^{-1}\beta^{-1}\sim\sigma_1^{w;2,1}.\]
Hence, braiding operations associated to $\sigma_1^{v;2,1}$ and $\sigma_1^{w;2,1}$ are identical. This reasoning extends {\it mutatis mutandis} to situations when we have more than one pair of Majorana fermions on a $3$-connected network $\Gamma$. Namely fix a rooted spanning tree of the network, $T\subset \Gamma$, and pick any two essential vertices of the network that are connected by a $\Theta$-graph as in Fig.~\ref{theta-init}b). Denote such a $\Theta$-subgraph by $\Theta_{v,w}$. Consider simple braid $\sigma_j^{v;\ba,2,1}$ (as $\Gamma$ is $2$-connected, the choice of $\ba$ does not matter). We need to make the following additional assumptions: i) $j$ is odd, ii) the root $*\in T\subset \Gamma$ is disjoint with $\Theta_{v,w}$ and iii) $\Gamma$ is large enough so that one can accommodate anyons $1,\hdots,j-1$ outside $\Theta_{v,w}$ 'over' vertex $w$ (in the sense of vertex order imposed by the choice of the spanning tree $T$) and keep them fixed during the exchange. Then, we are set out to repeat the reasoning for a single pair of Majorana fermions exchanging on a $\Theta$-graph to obtain that $\sigma_j^{v;\ba,2,1}\sim \sigma_j^{w;\ba',2,1}$ up to one-particle moves for any choice of $\ba$ and $\ba'$. Unfortunately, it is not clear to us how to extend this methodology to simple braids with even $j$.


\begin{thebibliography}{99}
\bibitem{reviews} C. Nayak, S. H. Simon, A. Stern, M. Freedman, and S. Das Sarma, Non-Abelian anyons and topological quantum computation, Rev. Mod. Phys. 80, 1083 (2008)
\bibitem{alicea-rev} J. Alicea, New directions in the pursuit of Majorana fermions in solid state systems, Rep. Prog. Phys. 75 076501 (2012)
\bibitem{alicea} J. Alicea, Y. Oreg, G. Refael, F. von Oppen and M. P. A. Fisher, Non-Abelian statistics and topological quantum information processing in 1D wire networks, Nature Physics 7, 412–417 (2011)
\bibitem{BM} Byung Hee An, T. Maciazek, Geometric presentations of braid groups for particles on a graph, arXiv:2006.15256 (2020)
\bibitem{al-in-as} A. Das, Y. Ronen, Y. Most, Y. Oreg, M. Heiblum and H. Shtrikman, Zero-bias peaks and splitting in an Al–InAs nanowire topological superconductor as a signature of Majorana fermions, Nature Physics 8, 887–895 (2012)
\bibitem{mourik} V. Mourik, K. Zuo, S. M. Frolov, S. R. Plissard, E. P. A. M. Bakkers, L. P. Kouwenhoven, Signatures of Majorana Fermions in Hybrid Superconductor-Semiconductor Nanowire Devices, Science 336 (6084), 1003-1007 (2012)
\bibitem{rokhison} L. P. Rokhinson, X. Liu and J. K. Furdyna, The fractional a.c. Josephson effect in a semiconductor–superconductor nanowire as a signature of Majorana particles, Nature Physics 8, 795–799 (2012)
\bibitem{perge} S. Nadj-Perge, I. K. Drozdov, J. Li, H. Chen, S. Jeon, J. Seo, A. H. MacDonald, B. A. Bernevig, A. Yazdani, Topological matter. Observation of Majorana fermions in ferromagnetic atomic chains on a superconductor, Science 346 (6209), 602-607 (2014)
\bibitem{pachos} Jin-Shi Xu, K. Sun, Y.-J. Han, Chuan-Feng Li, J. K. Pachos anf Guang-Can Guo, Simulating the exchange of Majorana zero modes with a photonic system, Nature Communications 7, 13194 (2016) 
\bibitem{flux} T. Hyart, B. van Heck, I. C. Fulga, M. Burrello, A. R. Akhmerov, C. W. J. Beenakker, Flux-controlled quantum computation with Majorana fermions, Phys. Rev. B 88, 035121 (2013)
\bibitem{rozek} T. E. O'Brien, P. Ro\.{z}ek, A. R. Akhmerov, Majorana-based fermionic quantum computation, Phys. Rev. Lett. 120, 220504 (2018)
\bibitem{josephson1} F. Pientka, A. Keselman, E. Berg, A. Yacoby, A. Stern, and B. I. Halperin, Topological Superconductivity in a Planar Josephson Junction, Phys. Rev. X 7, 021032 (2017)
\bibitem{josephson2} A. Stern and E. Berg, Fractional Josephson Vortices and Braiding of Majorana Zero Modes in Planar Superconductor-Semiconductor Heterostructures, Phys. Rev. Lett. 122, 107701 (2019)
\bibitem{LM} J. M. Leinaas, J. Myrheim, On the theory of identical particles, Nuovo Cim. 37B, 1-23 (1977)
\bibitem{Souriau} J. M. Souriau, Structure des systmes dynamiques, Dunod, Paris (1970)
\bibitem{Einarsson} T. Einarsson, Fractional statistics on a torus, Phys. Rev. Lett. 64 (1995)
\bibitem{Wilczek} F. Wilczek, Fractional statistics and anyon superconductivity, Singapore: World Scientific (1990)
\bibitem{braids} K. Murasugi, B. Kurpita, A Study of Braids, Mathematics and Its Applications, Vol. 484, Springer Netherlands (1999)
\bibitem{menger} Karl Menger, Zur allgemeinen Kurventheorie, Fundamenta Mathematicae 10 (1), 96-115 (1927)
\bibitem{HKR} J. M. Harrison , J. P. Keating  and J. M. Robbins, Quantum statistics on graphs, Proc. R. Soc. A 467 (2125), 212–23 (2011)
\bibitem{HKRS} J. M. Harrison, J. P. Keating, J. M. Robbins and A. Sawicki, n -Particle Quantum Statistics on Graphs, Comm. Math. Phys. 330 (3), 1293-1326 (2014)
\bibitem{MS16} T. Maci\k{a}\.zek and A. Sawicki, Homology groups for particles on one-connected graphs, J. Math. Phys. 58 (6), 062103 (2017)
\bibitem{MS18} T. Maci\k{a}\.zek and A. Sawicki, Non-abelian quantum statistics on graphs, Comm. Math. Phys. 371 (3), 921-973 (2019)
\bibitem{KoPark} K. H. Ko and H. W. Park, Characteristics of graph braid groups, Discrete \& Computational Geometry, 48 (4), 915-963 (2012)
\bibitem{oppen-notes} F. von Oppen, Y. Peng, F. Pientka, Topological superconducting phases in one dimension, Topological Aspects of Condensed Matter Physics: Lecture Notes of the Les Houches Summer School 103 (2014)
\bibitem{kitayev} A. Yu. Kitaev, Unpaired Majorana fermions in quantum wires, Physics-Uspekhi 44, Supplement (2001)
\end{thebibliography}
\end{document}